# Understanding Dynamics in Coarse-Grained Models: III. Roles of Rotational Motion and Translation-Rotation Coupling in Coarse-Grained Dynamics

Jaehyeok Jin,[1,2] Eok Kyun Lee,[3] and Gregory A. Voth[1*]

[1] Department of Chemistry, Chicago Center for Theoretical Chemistry, Institute for Biophysical Dynamics, and James Franck Institute, The University of Chicago, Chicago, IL 60637, USA
[2] Department of Chemistry, Columbia University, New York, NY 10027, USA
[3] Department of Chemistry, Korea Advanced Institute of Science and Technology (KAIST), Daejeon 34141, South Korea

* Corresponding author: gavoth@uchicago.edu

**Abstract**
This paper series aims to establish a complete correspondence between fine-grained (FG) and coarse-grained (CG) dynamics by way of excess entropy scaling (introduced in Paper I). While Paper II successfully captured translational motions in CG systems using a hard sphere mapping, the absence of rotational motions in single-site CG models introduces differences between FG and CG dynamics. In this third paper, our objective is to faithfully recover atomistic diffusion coefficients from CG dynamics by incorporating rotational dynamics. By extracting FG rotational diffusion, we unravel, for the first time reported to our knowledge, a universality in excess entropy scaling between the rotational and translational diffusion. Once the missing rotational dynamics are integrated into the CG translational dynamics, an effective translation-rotation coupling becomes essential. We propose two different approaches for estimating this coupling parameter: the rough hard sphere theory with acentric factor (temperature-independent) or the rough Lennard-Jones model with CG attractions (temperature-dependent). Altogether, we demonstrate that FG diffusion coefficients can be recovered from CG diffusion coefficients by (1) incorporating "entropy-free" rotational diffusion with translation-rotation coupling and (2) recapturing the missing entropy. Our findings shed light on the fundamental relationship between FG and CG dynamics in molecular fluids.

# I. Introduction

This paper is the third in a series (with the preceding articles referred to as Paper I[1] and II[2] hereafter) concerning the artificially "fast" dynamics in coarse-grained (CG) systems.[3] In order to bridge the accelerated CG dynamics with respect to the reference fine-grained (FG, i.e., usually atomistic-level) dynamics, we extensively employed the excess entropy scaling formalism that explicitly links a thermodynamic property of the system to its dynamics. These efforts originated from Rosenfeld's scaling relationship:[4-6]

$$D^* = D_0 \exp(\alpha s_{ex}),  \quad (1)$$

where $D^*$ is the dimensionless diffusion coefficient rescaled by macroscopic units

$$D^* = D \frac{\rho^{\frac{1}{3}}}{\left(\frac{k_B T}{m}\right)^{\frac{1}{2}}}, \quad (2)$$

and the molar excess entropy is defined as

$$s_{ex} = \frac{S_{ex}}{N k_B} = \frac{1}{N k_B} \left( S(\rho, T) - S_{id}(\rho, T) \right). \quad (3)$$

In Paper I,[1] we developed a generalized theory to account for various modal contributions in order to compute excess entropy. We particularly focused on water by utilizing our recently developed CG model: BUMPer (Bottom-Up Many-Body Projected Water).[7, 8] For FG and CG water systems, we discovered that Eq. (1) holds with a similar exponent $\alpha$ for the same molecular system upon the coarse-graining process:

$$\ln D^*_{FG} = \alpha^{FG} s^{FG}_{ex} + \ln D^{FG}_0 = 0.73 \times s^{FG}_{ex} + 2.15, \quad (4)$$

$$\ln D^*_{CG} = \alpha^{CG} s^{CG}_{ex} + \ln D^{CG}_0 = 0.70 \times s^{CG}_{ex} - 0.35. \quad (5)$$

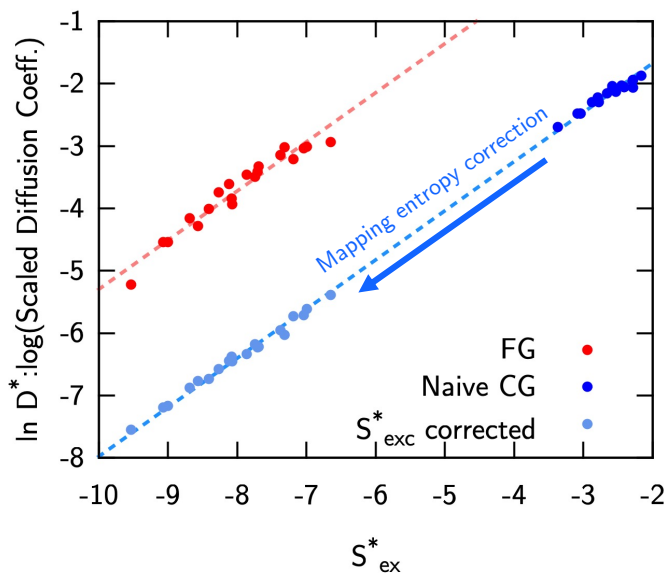

**Figure 1:** Excess entropy scaling for the FG (red dot) and CG (blue dot) water systems. We also plot the scaling relationships discovered in Paper I:[1] FG relationship [dashed red line, Eq. (4)] and CG relationship [dashed blue line, Eq. (5)]. To match FG and CG dynamics, the excess entropy is corrected for the CG system as shown in sky blue dot, but it does not agree with the FG dynamics due to the differences in $D_0$.

Our ultimate goal is to establish a complete correspondence between FG and CG dynamics, as shown in Fig. 1. From Eqs. (4) and (5), the characteristic terms that differentiate FG and CG dynamics stem from the intercept, or the "entropy-free" diffusion coefficient $D_0$, and the excess entropy $s_{ex}$. The differences in excess entropy between the FG and CG systems are referred to as the "missing entropy" (or mapping entropy).[9,10] We recently discovered that the missing entropies are entropic contributions from motions "beneath" the resolution of CG models that are lost during the coarse-graining process.[10]

Even though we can estimate the missing entropy term, $s_{ex}$, to recover the FG dynamics from the CG model, as depicted in Fig. 1, the entropy-free diffusion coefficient $D_0$ is relatively difficult to interpret, as this quantity does not have any clear physical origin. However, in Paper II, we developed a theory to analytically derive the $D_0$ term for CG models.[2] The central idea of Paper II was twofold: (i) Single-site CG models have the same resolution as hard sphere liquids and (ii) repulsive interactions in short-range regions mostly determine the structure of CG systems (from classical perturbation theory[11-16]). Therefore, we developed a new theory to construct an equivalent hard sphere system by preserving the long wavelength density fluctuations[17-19] of the given CG system, which we named "fluctuation matching". Since the hard sphere dynamical properties are expressed as a function of effective packing fraction, or effective diameter, such a hard sphere mapping allows for determining the entropy-free dynamical properties as well.[20-23] The fluctuation matching approach provides an analytical formulation for $D_0$ that recapitulates the trends from the CG simulations. Also, we were able to predict the accelerated CG diffusion coefficients at different temperatures based on FG information only. Nevertheless, this analytical theory is not directly applicable to FG systems, as various motions (not only translational) exhibited at the atomistic resolution are correlated. The present paper is devoted to filling this gap by presenting a more complete dynamical correspondence between the CG and FG systems.

The central differences between the single-site CG model and the reference FG model are ascribed to rotational motions exhibited at the FG resolution. Although FG particles involving in rotational motions are integrated out at the center-of-mass (CG) level, an effect of rotation on overall FG translation must be considered as the rotational diffusion occurs concurrently with translational diffusion during FG simulations.[24-26] In contrast, a single-site CG model can only have translational motions as rotational motions are integrated out with the coarse-graining process. Since introducing rotations to the system will slow down the overall dynamics by more rapidly decaying the velocity autocorrelations, the accelerated CG dynamics can be effectively corrected by recovering the missing dynamical features that originate from the FG behavior. However, two important questions arise: (1) How to extract the missing rotational motions from the FG system, and (2) how to introduce and couple the extracted rotational diffusion to the existing translational diffusion of the CG system?

First, applying the excess entropy scaling proposed by Rosenfeld to rotational motions is not straightforward because the choice of scaling schemes and excess entropies is ambiguous.

Furthermore, even if one establishes the correct scaling relationship for rotational diffusion that is consistent with the original Rosenfeld scheme, it is unclear how to link rotational diffusion with translational diffusion because their units are different. In this paper, we resolve this conundrum by projecting rotational displacements onto translational components based on our previous findings in which CG sites can be regarded as effective hard spheres. This so-called "arc approximation" is expected to effectively project rotational displacements on the hard sphere surface to translational displacements, allowing for the Rosenfeld scaling to be employed to rotational motions. An effort to link translational diffusion with rotational diffusion is not completely new as one can relate classical hydrodynamic theory to simple kinetic theory. In such a way, it is shown that translational diffusion described by the Stokes-Einstein relation[27] is explicitly related to the Stokes-Einstein-Debye relation[28] that determines rotational diffusion. Nevertheless, our attempt to assess rotational diffusion as missing motions from the coarse-graining process is an alternative and new approach. Excess entropy terms relevant to rotational diffusion can be readily obtained from our framework demonstrated in Paper I, as well as theories developed by Lazaridis, Karplus,[29] and Zielkiewicz.[30] When the Rosenfeld-like scaling scheme is in place, a natural extension is to also examine if the same scaling law will hold for rotational diffusion. This validation will serve as proof of universality in excess scaling relationships for different underlying motions.

In turn, the missing dynamics in CG simulations can be recuperated by including projected rotational components from the FG point of view. However, once we reintroduce the rotational degrees of freedom to the translational CG motions, exchanges between angular and linear momentum upon collision occur, and the so-called "translation-rotation coupling" must be considered.[31-36] In Paper II, CG diffusion was described as an effective hard sphere diffusion from the Enskog theory,[2] which provides accurate dynamic properties over various density ranges but is limited in that it was developed for perfectly "smooth hard spheres" in which the particles do not exchange any angular momentum upon collision.[37, 38] This assumption works well for the single-site CG model where there is only translational motion – there would be no need to consider angular momentum exchanges. However, if we integrate the FG rotational diffusion to the CG particles, we must account for the angular momentum exchanges upon collision.

To correctly address the coupled motions in such CG models, we adopt the concept of "rough hard spheres" introduced by Chandler[39] and extensively developed in the 1970-80s.[39-43] To analytically model the reduced diffusion from rapidly decaying velocity autocorrelations under translation-rotation coupling, Chandler developed a rough hard sphere theory where the diffusion coefficient is a smooth hard sphere diffusion coefficient scaled by a roughness parameter, which accounts for the coupling between translational and rotational motions upon collision.[39, 42] Based on this theory, we assert that the full FG dynamics can be recovered from the CG dynamics with a correct translation-rotation coupling once the roughness parameter of water is determined. Since the roughness parameter is highly related to the non-sphericity of molecules, we leverage the reported correlation between the roughness parameter and the non-sphericity of molecular liquids[44] to derive the translation-rotation coupling factor for water.

In summary, we study here the aforementioned connections between FG and CG dynamics and aim to recover the FG diffusion coefficients of water from the simple CG diffusion coefficients at different temperatures.

## II. Theory
### 2-1. Rotational Diffusion at FG Resolution

In Paper I, we obtained the translational diffusion coefficient by employing Einstein's relation to the center-of-mass mean squared displacement (MSD)

$$D_t = \lim_{t\to\infty} \frac{1}{6t}\langle R^2(t)\rangle = \lim_{t\to\infty} \frac{1}{6t}\frac{1}{N_{CG}}\sum_{I}^{N_{CG}}\left|\vec{R}_I(t) - \vec{R}_I(0)\right|^2, \quad (6)$$

where $\mathbf{R}_I(t)$ denotes the position of the center-of-mass of molecule $I$ at time $t$. Analogously, the rotational diffusion coefficient is defined as

$$D_{\text{rot}} = \lim_{t\to\infty} \frac{1}{4t}\langle \phi^2(t)\rangle = \lim_{t\to\infty} \frac{1}{4t}\frac{1}{N_{CG}}\sum_{I}^{N_{CG}}\left|\vec{\phi}_I(t) - \vec{\phi}_I(0)\right|^2. \quad (7)$$

In Eq. (7), $\langle \phi^2(t)\rangle$ is the rotational mean square displacement represented as a rotational displacement vector $\vec{\phi}_I(t)$ for the molecule $I$ at time $t$. However, unlike translational motions, the rotational displacement vector $\vec{\phi}_I(t)$ by definition is bounded from 0 to $2\pi$ rad. Hence, to assure the unboundedness of $\vec{\phi}_I(t)$, we alternatively divide the changes in rotational displacements as discrete vectors and integrate them up at each step. Figure 2 demonstrates a schematic procedure for constructing $\vec{\phi}_I(t)$ from the normalized polarization vector $\hat{p}_I(t)$ following the original work.[45]

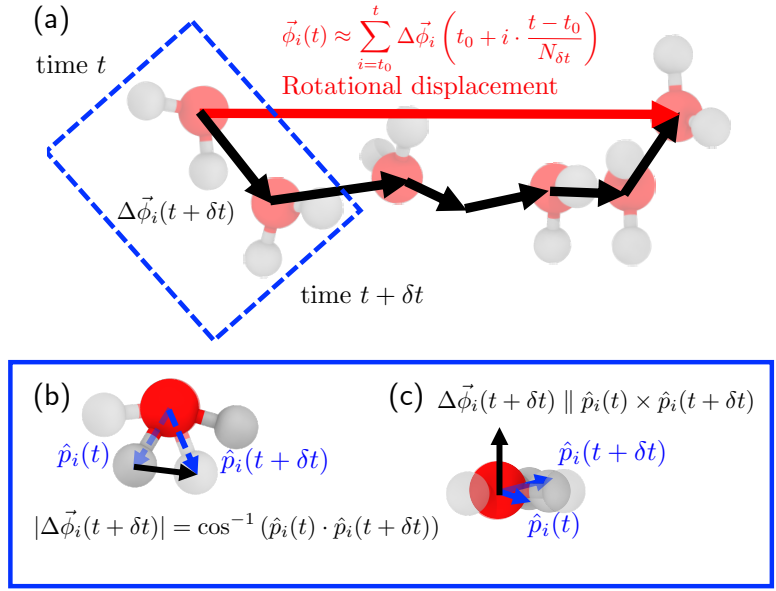

**Figure 2.** Schematic procedure for calculating the unbounded rotational mean square displacements: (a) FG trajectory of rotational motions for water from (b) top-view and (c) side-view. For each molecule, we calculate the normalized polarization vector $\hat{p}_i$ (blue) at every time step. The time differential of the rotational displacement $\Delta\vec{\phi}$ (black arrow) is calculated using two

adjacent normalized polarization vectors at time $t$ and $t + \delta t$. Finally, the rotational displacement $\vec{\phi}$ is obtained from a summation of $\Delta\vec{\phi}$ as shown by the red line in (a).

The normalized polarization vector is defined as a vector from the center-of-mass position to the midpoint between two oxygen atoms, which is an axis for the $C_{2v}$ symmetry. For the two adjacent FG configurations at a time step of $t$ and $t + \delta t$ from the given FG trajectory, we define a differential of rotational displacement vector as $\Delta\vec{\phi}_I(t + \delta t)$ where the magnitude of the differential is an angle spanned by the polarization vectors from $t$ to $t + \delta t$

$$\left|\Delta\vec{\phi}_I(t + \delta t)\right| \coloneqq \cos^{-1}\left(\hat{p}_I(t) \cdot \hat{p}_I(t + \delta t)\right),\tag{8}$$

with direction given by

$$\Delta\vec{\phi}_I(t + \delta t) \parallel \hat{p}_I(t) \times \hat{p}_I(t + \delta t).\tag{9}$$

Therefore, we construct the differential vectors from adjacent configurations and perform a discrete integration to obtain the rotational displacement vector from the initial time $t_0$ to the final time $t$ with $N_{\delta t}$ steps

$$\vec{\phi}_I(t) \approx \sum_{i=t_0}^{t} \Delta\vec{\phi}_I\left(t + i \cdot \frac{t - t_0}{N_{\delta t}}\right).\tag{10}$$

By design, Equation (10) is no longer bounded, and we can utilize Eq. (7) to obtain $D_{\text{rot}}$.

**2-2. Excess Entropy Scaling for Rotational Diffusion: Challenges**
From the rotational diffusion coefficients obtained at the FG resolution, a Rosenfeld-like excess entropy scaling relationship can be designed as

$$D_{\text{rot}}^* = D_0^{\text{rot}} \exp(\alpha^{\text{rot}} s_{ex}^{\text{rot}}),\tag{11}$$

where $D_{\text{rot}}^*$ is the rescaled rotational diffusion, and $\alpha^{\text{rot}}$ is the scaling factor for rotational diffusion. To examine if Eq. (11) holds, two variables must be determined: $D_{\text{rot}}^*$ and $s_{ex}^{\text{rot}}$.

Yet, applying the excess entropy scaling for the rotational diffusion can be problematic since there are no macroscopic units that are consistent with the dynamics at the macroscopic level. For translational diffusion, it is shown that the rescaling scheme given by Eq. (2) is consistent with the characteristic length and timescales of Newtonian dynamics, and this can be further extended to the hidden scale invariance of strongly correlated systems.[46, 47] However, such a rescaling scheme does not exist in the case of rotational diffusion. To our knowledge, only a relatively small number of studies have been conducted on rotational diffusion scaling compared to that of translational diffusion. Most of these contributions from Chopra, Truskett, and Errington applied the following scaling scheme:[48-51]

$$D_{\text{rot}}^* = D_{\text{rot}} \frac{1}{\left(\frac{k_B T}{m}\right)^{\frac{1}{2}} \left(\frac{\rho}{m}\right)^{\frac{1}{3}}}.\tag{12}$$

Even though the above scaling scheme is suggested to demonstrate a Rosenfeld-style scaling, there is no apparent physical connection between the scaling scheme in Eq. (12) and the macroscopic units. The authors of Refs. 48-51 report that excess entropy scaling is observed in rotational diffusion using Eq. (12), but any scaling scheme will eventually give rise to a linear relationship since $\left(\frac{k_B T}{m}\right)$ and $\left(\frac{\rho}{m}\right)$ are constant. Therefore, a well-founded, physical scaling relationship is still needed for rotational dynamics.

Also, directly scaling rotational diffusion can be problematic as it lacks a physical link with translational diffusion. Note that rotational diffusion is in units of $\text{rad}^2 \cdot \text{s}^{-1}$, whereas translational diffusion is in units of $\text{m}^2 \cdot \text{s}^{-1}$. Therefore, the two diffusion coefficients are not directly related, while their dynamical behaviors must be somehow related. In order to quantitatively estimate the effect of rotational dynamics on the mapped FG translation, an alternative approach is needed to bridge between two different diffusion processes.

## 2-4. Arc Approximation: Link Rotation to Translation

In Paper II, we demonstrated that it is physically reasonable to treat the single-site CG site as a hard sphere with an effective diameter that gives consistent dynamics with the CG reference.[2] Hence, at the mapped FG resolution, the overall FG trajectories can be thought of as hard sphere trajectories containing both rotation and translation. Since the effective hard sphere diameter for a given FG trajectory remains unchanged, the rotational displacement would still not deviate from the hard sphere.

In this regard, translational displacement arising from rotational motions can be projected onto the hard sphere surface. Then, under the hard sphere approximation (thus no changes in hard sphere diameters), it is reasonable to approximate translational components due to rotations as the arc length $\sigma_{\text{t-rot}}$ on the hard sphere surface with its diameter $R_{\text{HS}}$

$$\sigma_{\text{t-rot}} \cong R_{\text{HS}} \cdot \phi_I(t), \quad (13)$$

where $\phi_I(t) = |\vec{\phi}_I(t)|$. We will provide a detailed discussion for computing $R_{\text{HS}}$ for a given CG system in Section 3-2. Equation (13) directly links rotational displacements to translational displacements using a hard sphere approximation, which we denote as the arc approximation. By adopting this approach, the effective translational diffusion coefficient due to rotation, as denoted by $D_{\text{t-rot}}$, can be formulated in terms of Einstein's relation

$$D_{\text{t-rot}} = \lim_{t \to \infty} \frac{1}{6t} \langle \sigma_{\text{t-rot}}^2(t) \rangle = \lim_{t \to \infty} \frac{1}{6t} \frac{1}{N_{CG}} \sum_{I}^{N_{CG}} |R_{HS} \cdot \phi_I(t) - R_{HS} \cdot \phi_I(0)|^2. \quad (14)$$

We note that a similar description for the anisotropic diffusion coefficients of colloidal particles has been suggested as the product between the distance from the center and rotational diffusion motions.[52, 53] Another physical interpretation underlying Eq. (13) is possible at the hydrodynamic level, known as the Stokes-Einstein[54, 55] and Stokes-Einstein-Debye[28] relationships. An extension of the present approach to hydrodynamic description will be discussed in a subsequent paper in this series.

In turn, we can reapply the original Rosenfeld scaling to the translational displacements resulting from rotational diffusion as

$$D_{\text{t-rot}}^* = D_0^{\text{t-rot}} \exp(\alpha^{\text{t-rot}} s_{ex}^{\text{t-rot}}).$$

(15)

Since $D_{\text{t-rot}}$ denotes the translational diffusion in units of m²·s⁻¹, we can apply the same rescaling scheme from the original work by Rosenfeld to obtain the dimensionless diffusion coefficient

$$D^*_{\text{t-rot}} = D_{\text{t-rot}} \frac{\rho^{\frac{1}{3}}}{\left(\frac{k_B T}{m}\right)^{\frac{1}{2}}}.$$

(16)

The last term to be determined before employing Eq. (15) is the excess entropy term related to rotation, $s_{ex}^{\text{t-rot}}$.

### 2-3. Excess Entropy for Rotation

In Paper I, we proposed a calculation scheme for excess entropy[1] based on the two-body contribution from the multiparticle correlation expansion[56-59]

$$S^{(2)} = -2\pi \int_0^\infty \{g^{(2)}(\mathbf{r}) \ln g^{(2)}(\mathbf{r}) - [g^{(2)}(\mathbf{r}) - 1]\} \mathbf{r}^2 \cdot d\mathbf{r},$$

(17)

in which $g^{(2)}(r)$ is the pair distribution function. Since pair $\mathbf{r}$ in Eq. (17) includes both the pair distance $r = |\mathbf{r}| = |\mathbf{r}_2 - \mathbf{r}_1|$ and the orientation of each particle $\omega_1$ and $\omega_2$, an accurate pair excess entropy was obtained by considering contributions from both translations and orientations using the decomposition method suggested by Lazaridis, Karplus,[29] and Zielkiewicz[30]

$$g^{(2)}(\mathbf{r}_{12}) = g^{(2)}(r, \omega_1, \omega_2) = g^{(2)}_{\text{trans}}(r) \cdot g^{(2)}_{\text{or}}(\omega_1, \omega_2 | r).$$

(18)

We have previously shown that this decomposition approach shares an identical factorization scheme with the two-phase thermodynamic (2PT) method:[60-62]

$$S^{\text{FG}} = S^{\text{FG}}_{\text{trn}} + S^{\text{FG}}_{\text{or}} = S^{\text{FG}}_{\text{trn}} + S^{\text{FG}}_{\text{rot}} + S^{\text{FG}}_{\text{vib}}.$$

(19)

Since contributions from vibrational motions to excess entropy are negligible in water, the rotational entropies from Eq. (19) should correspond to the orientational entropy from the definition given by Lazaridis and Karplus.[29] In order to apply excess entropy scaling for rotation, we followed Refs. 48-51 and chose the excess entropy associated with the FG rotational motion to be the overall excess entropy of the FG system, i.e.,

$$s_{ex}^{\text{t-rot}} = s_{ex}^{\text{FG}}.$$

(20)

Equation (20) can be further understood as follows. Since the computed rotational diffusion coefficients are obtained from the FG trajectories involving both translation and rotation at the same time, the extracted rotational motion is not completely independent from the center-of-mass translation, as it is conditional to its translational motion: $\phi_I(t) = \phi_I(t|R(t))$.

Finally, the Rosenfeld-like excess entropy scaling for translational diffusion due to the projected rotational diffusion is formulated as

$$D^*_{\text{t-rot}} = D_{\text{t-rot}} \cdot \frac{\rho^{\frac{1}{3}}}{\left(\frac{k_B T}{m}\right)^{\frac{1}{2}}} = D_0^{\text{t-rot}} \exp(\alpha^{\text{t-rot}} s_{ex}^{\text{FG}}).$$

(21)

Our main objective here is to check if Eq. (21) holds, and, if so, to compare the fitted exponent $\alpha^{\text{t-rot}}$ with the exponents from translational diffusion that are universal to FG and CG trajectories $\alpha^{\text{trans}} = \alpha^{\text{FG}} = \alpha^{\text{CG}}$.

### 2-5. Translation-Rotation Coupling
From Chandler's work on translation-rotation coupling,[39] the diffusion coefficient of rough hard spheres, $D_{\text{RHS}}$, can be related to that of smooth hard spheres, $D_{\text{SHS}}$, as
$$D_{\text{RHS}} = A_D D_{\text{SHS}}. \tag{22}$$
The roughness parameter in Eq. (22), $A_D$, is bounded from zero to unity, reflecting the translation-rotation coupling. Namely, $A_D$ is unity for perfectly smooth hard spheres and deviates to zero as a molecule becomes highly non-spherical. Thus, introducing a concept of rough hard spheres can effectively account for the translation-rotation coupling.

Even though the physical picture of $A_D$ is clear as a measure of sphericity of the given molecule, Chandler did not derive any systematic expressions for $A_D$ in his original work for carbon tetrachloride, reporting an $A_D = 0.54$ by comparing the Enskog-based smooth hard sphere diffusion coefficients to experiment.[63] After a decline in popularity of the hard sphere theory, the roughness parameters based on the hard sphere prediction and experiments were reported for only a handful of liquid systems in literature,[42] resulting in a lack of extensibility over other types of liquids. Despite its semi-empirical nature, after almost 30 years, Ruckenstein and Liu proposed an alternative way to assess the roughness parameter based on the non-sphericity of molecules.[44] Among various metrics for non-sphericity, they chose acentric factor $\omega$ that is known as a quantitative measure of the non-sphericity based on
$$\omega = -\log P_{vp,r}(T_r = 0.7) - 1, \tag{23}$$
where $P_{vp,r}$ is the reduced vapor pressure $P_{vp,r} = P/P_c$ at the reduced temperature $T_r = T/T_c = 0.7$.[64, 65] Pitzer found that non-spherical liquids deviate from the ideal behavior described by the theorem of the corresponding states[66] in the pressure-temperature phase diagram, and this deviation can be quantified as Eq. (23). Since the acentric factor represents the non-sphericity of a molecule and can also be measured in experiment, it has become a standard factor for phase characterization and for improving phase-behavior computations from equations of state. In this regard, the central argument of Ruckenstein and Liu was to establish a link between the roughness parameter and the acentric factor by examining their correlations over 42 different data sources for 26 different liquid systems.[44] In the end, they found a correlation between $A_D$ and $\omega$ as
$$A_D = 0.9673 - 0.2527\omega - 0.70\omega^2. \tag{24}$$
For unknown liquid molecules, we claim that Eq. (24) can yield a suitable translation-rotation coupling parameter by examining how the liquid molecule is non-spherical, and this idea can be applied for our CG water case. Yet, we also note that this correlation is somewhat *ad hoc*, and thus some errors may arise when applying Eq. (24) to systems beyond the parametrized sets, such as water.

### 2-6. Computational Details
The computational details and results of this paper are along the same lines as the previous papers in our series.[1, 2] In this subsection, we briefly provide essential information needed to describe the system setting in this work. Readers are referred to Paper I for the complete set-up of the FG and CG simulations and the CG parameterization.

In general, the bottom-up CG interaction parameters are derived from FG simulations to precisely reproduce structural correlations.[3, 67-75] Particularly, in this work, the CG model of water (BUMPer) is parameterized from four different FG force fields: SPC/E, SPC/Fw, TIP4P/2005, and TIP4P/Ice.[7, 8] Unlike conventional pairwise CG models, these BUMPer interactions are designed by projecting three-body Stillinger-Weber interactions onto pairwise basis sets, and thus the integrated three-body contributions give rise to high-fidelity CG models at inexpensive computational cost, resulting in a more faithful recapitulation of two-body and *N*-body correlations. In the earlier BUMPer work, we designed the BUMPer force fields to be temperature transferable, and thus the effective interactions can span from 280 K to 360 K for every 20 K at 1 atm (spans the entire liquid range at 1 atm).[8] We adopt this temperature range (five different temperatures) to analyze the effect of temperature on self-diffusion.

## III. Results
### 3-1. Rotational Diffusion in Water

Figure 3 depicts the rotational mean square displacements for four different BUMPer models parameterized by different FG force fields at different temperatures. The monotonically increasing feature of the rotational mean square displacements with time confirms the unboundedness of the devised scheme in Fig. 2. As expected, the rotational displacements also increase as temperature increases from 280 K to 360 K.

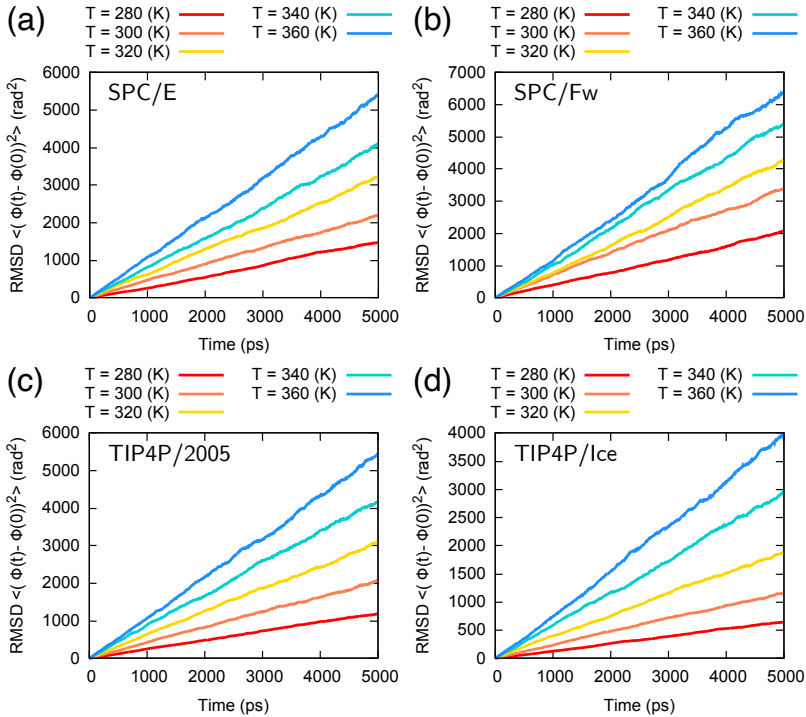

**Figure 3.** Rotational mean square displacement $\langle \phi^2(t) \rangle$ from the FG trajectories at temperatures from 280 K to 360 K: (a) SPC/E, (b) SPC/Fw, (c) TIP4P/2005, and (d) TIP4P/Ice. While the rotational displacement increase with temperature, its unit (rad$^2$) is not consistent with the translational displacement.

In Table 1(a), we calculate $D_\text{rot}$ by employing Eq. (7) to the scheme described in Fig. 2. The computed values are listed in a similar order as the previous work utilizing the SPC/E force fields, as $10^{-1}$-$10^{-2}$ rad$^2$·ps$^{-1}$ from Ref. 76 or $10^{10}$-$10^{11}$ rad$^2$·s$^{-1}$ from Ref. 49, confirming the validity of our calculations. However, as stated in Paper I,[1] translational diffusion occurs at different orders and has units of m$^2$·s$^{-1}$. (One interesting avenue to explore for future work is to link the Stokes-Einstein relation and Stokes-Einstein-Debye relation for water.[27, 28]) In turn, due to the differences in the units for translational and rotational diffusion, applying the Rosenfeld-like scaling schemes and recovering the missing dynamics are not possible, as illustrated in Fig. 3.

**Table 1:** Rotational diffusion coefficients of water evaluated for FG models: (a) SPC/Fw, (b) SPC/E, (c) TIP4P/2005, and (d) TIP4P/ice. We calculated the effective rotational diffusivity $D_\text{rot}$ and its translation component $D_\text{t-rot}$ by employing the arc approximation at various temperatures (280-360 K).

| (a) SPC/E | | | (b) SPC/Fw | | |
|---|---|---|---|---|---|
| Temperature | $D_\text{rot}$ (rad$^2$·ps$^{-1}$) | $D_\text{t-rot}$ (cm$^2$·s$^{-1}$) | Temperature | $D_\text{rot}$ (rad$^2$·ps$^{-1}$) | $D_\text{t-rot}$ (cm$^2$·s$^{-1}$) |
| 280 K | 7.65×10$^{-2}$ | 8.19×10$^{-6}$ | 280 K | 1.01×10$^{-1}$ | 1.08×10$^{-5}$ |
| 300 K | 1.08×10$^{-1}$ | 1.17×10$^{-5}$ | 300 K | 1.68×10$^{-1}$ | 1.82×10$^{-5}$ |
| 320 K | 1.58×10$^{-1}$ | 1.72×10$^{-5}$ | 320 K | 2.18×10$^{-1}$ | 2.37×10$^{-5}$ |
| 340 K | 2.02×10$^{-1}$ | 2.23×10$^{-5}$ | 340 K | 2.77×10$^{-1}$ | 3.05×10$^{-5}$ |
| 360 K | 2.69×10$^{-1}$ | 3.00×10$^{-5}$ | 360 K | 3.31×10$^{-1}$ | 3.69×10$^{-5}$ |
| (c) TIP4P/2005 | | | (d) TIP4P/Ice | | |
| Temperature | $D_\text{rot}$ (rad$^2$·ps$^{-1}$) | $D_\text{t-rot}$ (cm$^2$·s$^{-1}$) | Temperature | $D_\text{rot}$ (rad$^2$·ps$^{-1}$) | $D_\text{t-rot}$ (cm$^2$·s$^{-1}$) |
| 280 K | 6.00×10$^{-2}$ | 6.41×10$^{-6}$ | 280 K | 3.28×10$^{-2}$ | 3.53×10$^{-6}$ |
| 300 K | 1.02×10$^{-1}$ | 1.11×10$^{-5}$ | 300 K | 5.77×10$^{-2}$ | 6.26×10$^{-6}$ |
| 320 K | 1.52×10$^{-1}$ | 1.66×10$^{-5}$ | 320 K | 9.39×10$^{-2}$ | 1.03×10$^{-5}$ |
| 340 K | 2.09×10$^{-1}$ | 2.31×10$^{-5}$ | 340 K | 1.47×10$^{-1}$ | 1.64×10$^{-5}$ |
| 360 K | 2.71×10$^{-1}$ | 3.04×10$^{-5}$ | 360 K | 2.00×10$^{-1}$ | 2.25×10$^{-5}$ |

### 3-2. Excess Entropy Scaling of Rotational Diffusion

Introducing the arc approximation allows one to quantitatively assess the role of rotations in excess entropy scaling. In order to utilize Eq. (13), the effective hard sphere radius $R_{HS}$ must be determined beforehand. Among various hard sphere approximations,[77] Paper II discussed two different mapping approaches for CG models, the Barker-Henderson perturbation and fluctuation matching, to obtain an effective hard sphere diameter or packing fraction.[2]

In this section, we employ the Barker-Henderson approach to obtain the hard sphere property in order to employ the arc approximation [Eq. (13)]. Since the rotational motion is confined to each molecule, we are not interested in capturing the long wavelength properties, which are of particular interest in fluctuation matching. Namely, we aim to analyze the many-body CG PMF on the basis of pairwise interactions and determine its effective diameter to approximate the size of the hard spheres. The work by Barker and Henderson gives the effective hard sphere diameter for the CG system as

$$\sigma_{BH} = \int_0^{R_0} [1 - \exp(-\beta U(R))] \cdot dR, \tag{25}$$

where $R_0$ is determined to the shortest possible distance that gives zero CG interaction: $U(R_0) = 0$. Then, the effective hard sphere radius of CG system is estimated as $\sigma_{BH}/2$. Equation (25) can be readily applied to the many-body CG PMFs at different temperatures that were depicted in Fig. 2 of Paper I.[1] Paper II provides a comprehensive discussion of the theories for hard sphere mapping,[2] and Table A1 in Appendix A lists the estimated $\sigma_{BH}$ values for different water models. Next, the translational components of the rotational mean square displacement $\langle (R_{HS}\phi)^2(t) \rangle$ are shown in Fig. 4. The temperature-dependent trend of the projected displacements also resembles the trend from the pure rotational displacements in that they only differ by the hard sphere radius.

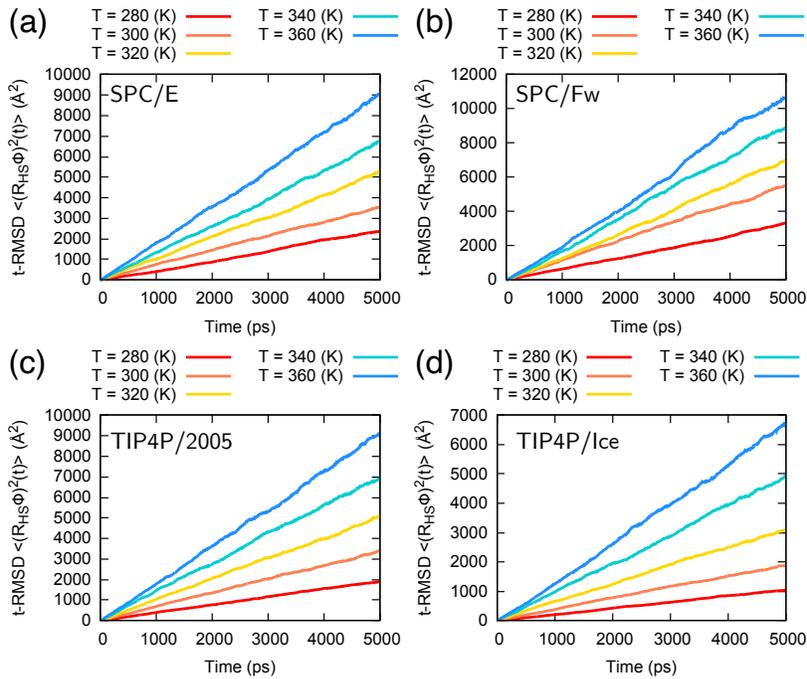

**Figure 4.** Translational-component of rotational mean square displacement $\langle (R_{HS}\phi)^2(t) \rangle$ from the FG trajectories at temperatures from 280 K to 360 K: (a) SPC/E, (b) SPC/Fw, (c) TIP4P/2005, and (d) TIP4P/Ice. The arc approximation allows for assessing the missing motions in terms of excess entropy scaling.

We now perform the excess entropy scaling to examine the scaling exponent for rotation. For four different FG force fields, Fig. 5(a) confirms that the scaling behavior follows a natural logarithm of the projected diffusion coefficients and is proportional to the overall FG excess entropies with the relationship

$$\ln D^*_{t-\text{rot}} = 0.73 \times s_{ex}^{FG} + 1.57. \tag{26}$$

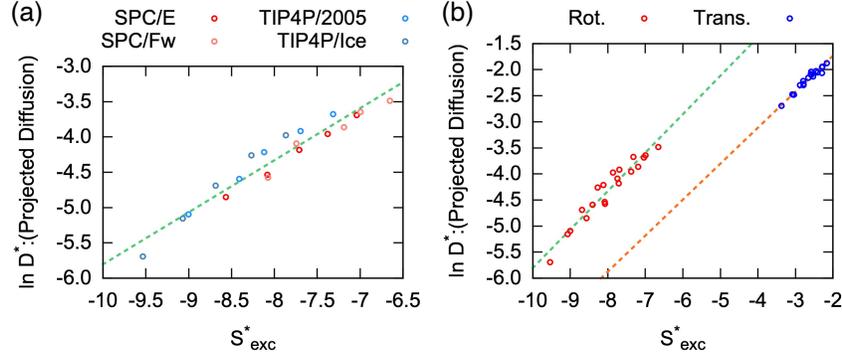

**Figure 5.** Examination of the Rosenfeld scaling for water rotation and its universality. (a) FG scaling relationship for the projected rotation where the SPC/E (red circles), SPC/Fw (green circles), TIP4P/2005 (blue circles), TIP4P/Ice (purple circles) models fall into a linear scaling relation shown in Eq. (26) (dashed green line). (b) Introduced excess entropy scaling schemes are invariant under different motions: rotation (red circles) and translation (blue circles). The rotational scaling [dashed green line, Eq. (26)] exhibits an almost identical slope as that of the translational scaling [dashed orange line, Eq. (5)], supporting the universality of the scaling relationships.

Remarkably, the central result of our work is that the scaling exponent from the projected rotation in Eq. (26) is almost identical to the scaling exponent from the translational motional in Eq. (5). Both equations are plotted in Fig. 5(b), indicating

$$\alpha^{\text{t-rot}} \approx \alpha^{\text{CG}} \approx \alpha^{\text{FG}}.$$

(27)

Given the fact that the excess entropy scaling of overall FG trajectory also gives almost a similar exponent $\alpha^{\text{FG}}$, our findings confirm the universality of the excess entropy scaling over different motions. This is particularly important because, from this universality, any differences in diffusion can be fully understood as changes in excess entropy and the entropy-free dynamics $D_0$. Furthermore, our findings provide a systematic framework that can be potentially useful to effectively describe the accelerated dynamics of any CG model beyond the single-site CG resolution. As long as the missing entropies and motions due to coarsening can be characterized, the present framework can predict such changes in the diffusion coefficients.

### 3-3. General Overview of Corrected CG Diffusion

Based on our findings in the previous section, we summarize our general understanding of the differences in FG and CG dynamics in this section (Fig. 6).

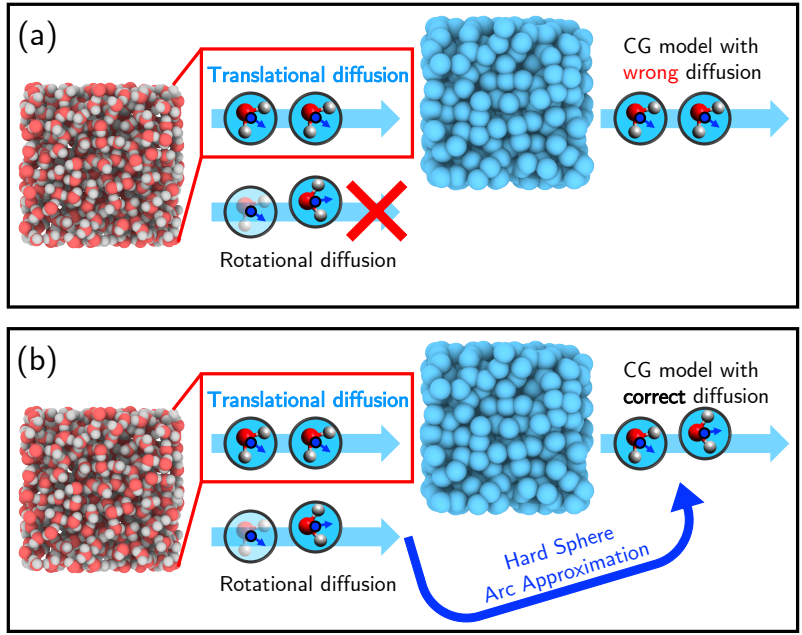

**Figure 6.** Schematic diagram for single-site CG dynamics of liquids. (a) Upon coarse-graining, the rotational degrees of freedom do not remain at the CG model, and thus the CG simulations represent only translational dynamics. (b) The correct representation of FG dynamics in the CG model can be done by integrating the rotational diffusion dynamics from the FG level onto the CG level. From a hard sphere point of view, we perform the arc approximation to assess the missing dynamics. Note that this process is done as a post-analysis of the naïve CG dynamics, not for constructing a new CG model.

By coarse-graining the liquid system to a single-site representation, the CG diffusion only contains translational motion since rotational motion does not remain at the CG resolution

$$D_{\text{CG}} = D^{\text{trans}} = D_0^{\text{trans}} \exp(\alpha s_{ex}^{\text{trans}}).$$

(28)

In Eq. (28), we proposed $\alpha$ as a universal exponent for all FG and CG systems. The parameterized CG model can only mimic translational motion since rotational diffusion is integrated out during the coarse-graining process. To correct for this accelerated CG dynamics, we introduce the rotational contributions from the FG level

$$D^{\text{t-rot}} = D_0^{\text{t-rot}} \exp(\alpha s_{ex}^{\text{FG}}).$$

(29)

Because Eq. (28) and (29) have different entropy terms $s_{ex}^{\text{trans}}$ and $s_{ex}^{\text{FG}}$, our approach explains why it is difficult to directly relate translational diffusion with rotational diffusion.

Applying this idea to water, Eq. (26) suggests that the missing rotational contribution to the reference FG diffusion is about $\ln D_0^{\text{t-rot}} = 1.57$. Since this contribution is completely missing in the CG description, we claim that this portion should be effectively added to the CG dynamics in order to recover the reference dynamics. However, this is also when the translation-rotation coupling comes into play.

### 3-3. Translation-Rotation Coupling of Water

We now introduce the idea of imposing a translation-rotation coupling. Our approach is different from the conventional hard sphere work by Chandler[39] as well as Berne and Montgomery.[33] From the conventional perspective, a hard sphere description was applied to accurately capture the overall diffusion behavior with respect to the reference diffusion. Thus, in such treatments, both translation and rotation should be considered at the same time while modeling the hard sphere system, resulting in a coupling between these two motions. However, our approach does not fully stem from hard sphere modeling as our primary purpose of using the hard sphere is to understand the CG translational diffusion.[2] In other words, we first constructed the CG model via bottom-up approaches, keeping only the translational degrees of freedom of the CG system. We do not have to consider translation-rotation coupling in the CG system, since there are no rotational degrees of freedom left. Therefore, by having only translational motion, we design an effective hard sphere system that can faithfully recapitulate translational motion, as seen from the CG simulation described in Paper II. However, once we integrate the rotational information from the FG trajectories to the resultant hard sphere system, both the translational and rotational motions appear in the hard sphere system at the CG resolution, and thus the resultant dynamics should be coupled.

In summary, the aforementioned procedures to recover the full dynamical information from CG diffusion can be performed in three sequential steps: (1) Correcting for $s_{ex}$ in Eq. (5), (2) incorporating the entropy-free rotational diffusion effect from $D_0^{t-rot}$, and (3) considering the translation-rotation coupling.

As initially introduced in Paper I and also depicted in Fig. 1 of this paper, correcting the excess entropy term by recuperating missing entropy is not enough to fully recover the FG dynamics because the rotational motions are not considered. By adding a translation-component of the rotational diffusion, the overall diffusion becomes more in agreement with the reference FG diffusion. Namely, by introducing the missing rotational motions, the overall CG dynamics effectively slows down compared to the pure CG translational diffusion. Finally, we consider the coupling between the rotational and translational diffusion. The coupling parameter, which is also the roughness parameter of water $A_D^{H_2O}$, is given by Eq. (24).

From the critical temperature and pressure measured for water in Ref. 78, the acentric factor can be estimated as $\omega_{H_2O} = 0.344$ using Eq. (23). As the acentric factor reflects the non-sphericity of a molecule, the computed value is within a reasonable range (less than 1). Using this value, we obtained the fitted roughness parameter $A_D^{H_2O} = 0.798$. That is, under angular and linear momentum exchanges in realistic collision conditions, the effective diffusion is expected to be reduced by ~ 20%.

### 3-5. Recovered Diffusion Coefficients: Temperature-Independent $A_D^{H_2O}$

We now quantitatively assess the recovered diffusion coefficient from the previous subsection in comparison to the reference diffusion coefficients from the FG simulations. Figure 7 provides an overview of the diffusion coefficients from the FG and CG simulations along with the recovered CG dynamics for comparison.

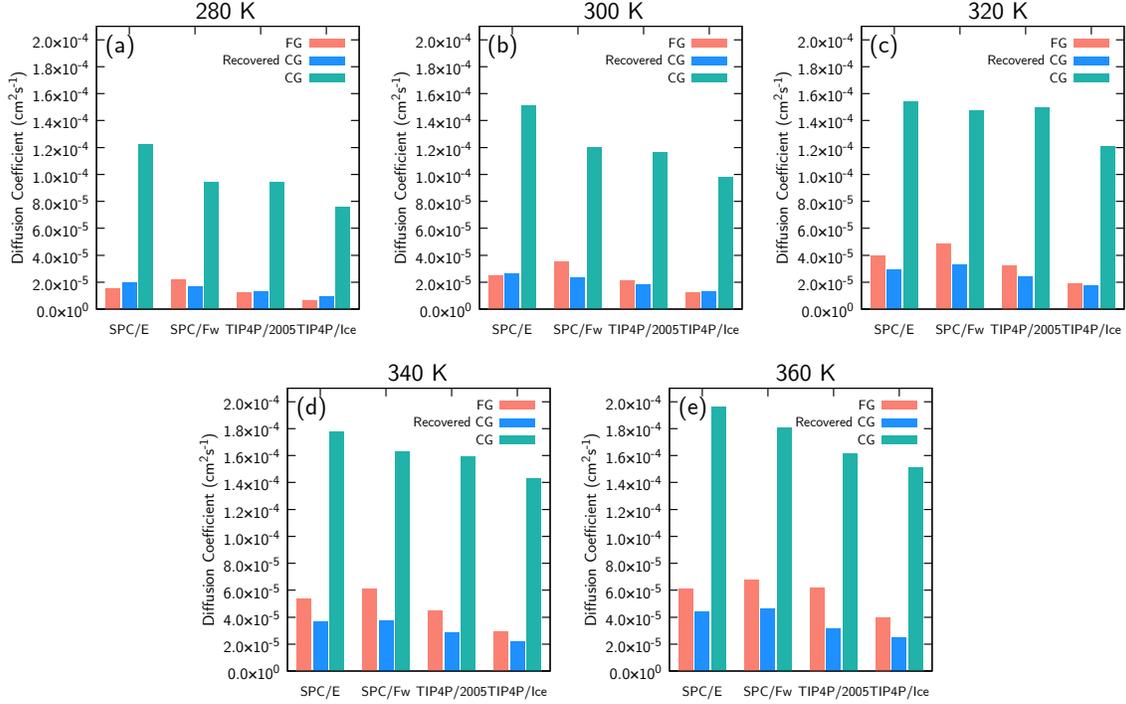

**Figure 7.** Diffusion coefficients of water for the FG (red solid) and CG (green solid) systems at different temperatures: (a) 280 K, (b) 300 K, (c) 320 K, (d) 340 K, and (e) 360 K. We recovered the FG dynamics (blue solid) from the CG diffusion coefficient values using a temperature-independent parameter $A_D$ [Eq. (24)].

Among the four different FG force fields, we find that SPC/Fw deviates the most by an average error of 32%. This difference can be understood from the nature of the force field: the oxygen-hydrogen bonds and hydrogen-oxygen-hydrogen angles are designed to be flexible unlike the rigid SPC/E or TIP4P-based models. Despite the fact that the vibrational modes of water are mostly negligible, this slight contribution violates the entropy partition scheme of Eq. (19). Also, while extracting the rotational diffusion, the polarization vector $\hat{p}_I(t)$ may not be able to fully capture rotation due to perturbations that originate from the oxygen-hydrogen bond vibrations. These vibrational contributions may result in deviations in the recovered diffusion coefficients.

The recovered CG diffusion coefficient $D_{CG}^{corr}$ and the error ratio $D_{CG}^{corr}/D_{FG}$ are presented in Table 2 and B1, repspectively. Our treatment, which incorporates rotational effects and translation-rotation coupling, successfully captures 72.6% of the FG dynamics from the CG dynamics. It is important to note that this agreement was achieved by strictly following the original description for the translation-rotation coupling reported by Chandler. In other words, in Eq. (22), $A_D$ was assumed to be rigorously density and temperature independent.[39] Nonetheless, with the introduction of the rough hard sphere model, high-pressure NMR spin echo techniques have allowed for the examinination of the temperature and density dependence of the coupling parameter.[79] In spin echo experiments, the accurate determination of diffusion coefficents is achieved by adjusting the coil current in the presence of a pulsed field gradient.[80] These experimental studies have revealed that the roughness parameter varies with temperature for small liquids.[81, 82]

**Table 2:** Diffusion coefficients of water evaluated for FG and CG models. Based on Fig. 7, the recovered CG diffusion coefficients ($D_{\text{CG}}^{\text{corr}}$) are listed for comparison and show high agreement with the FG diffusion coefficient at low temperatures. We demonstrated the fidelity of our approach by testing at various temperatures (280-360 K) using the four FG force models and their corresponding BUMPer CG models: (a) SPC/E, (b) SPC/Fw, (c) TIP4P/2005, and (d) TIP4P/Ice.

| (a) SPC/E (Diffusion in cm$^2$·s$^{-1}$) | | | | (b) SPC/Fw (Diffusion in cm$^2$·s$^{-1}$) | | | |
|---|---|---|---|---|---|---|---|
| Temp | $D_{\text{FG}}$ | $D_{\text{CG}}$ | $D_{\text{CG}}^{\text{corr}}$ | Temp | $D_{\text{FG}}$ | $D_{\text{CG}}$ | $D_{\text{CG}}^{\text{corr}}$ |
| 280 K | 1.55×10$^{-5}$ | 1.22×10$^{-4}$ | 1.96×10$^{-5}$ | 280 K | 2.20×10$^{-5}$ | 9.39×10$^{-5}$ | 1.70×10$^{-5}$ |
| 300 K | 2.51×10$^{-5}$ | 1.51×10$^{-4}$ | 2.64×10$^{-5}$ | 300 K | 3.53×10$^{-5}$ | 1.20×10$^{-4}$ | 2.31×10$^{-5}$ |
| 320 K | 3.98×10$^{-5}$ | 1.54×10$^{-4}$ | 2.91×10$^{-5}$ | 320 K | 4.84×10$^{-5}$ | 1.47×10$^{-4}$ | 3.26×10$^{-5}$ |
| 340 K | 5.33×10$^{-5}$ | 1.78×10$^{-4}$ | 3.63×10$^{-5}$ | 340 K | 6.08×10$^{-5}$ | 1.63×10$^{-4}$ | 3.77×10$^{-5}$ |
| 360 K | 6.12×10$^{-5}$ | 1.96×10$^{-4}$ | 4.37×10$^{-5}$ | 360 K | 6.77×10$^{-5}$ | 1.81×10$^{-4}$ | 4.64×10$^{-5}$ |
| (c) TIP4P/2005 (Diffusion in cm$^2$·s$^{-1}$) | | | | (d) TIP4P/Ice (Diffusion in cm$^2$·s$^{-1}$) | | | |
| Temp | $D_{\text{FG}}$ | $D_{\text{CG}}$ | $D_{\text{CG}}^{\text{corr}}$ | Temp | $D_{\text{FG}}$ | $D_{\text{CG}}$ | $D_{\text{CG}}^{\text{corr}}$ |
| 280 K | 1.20×10$^{-5}$ | 9.41×10$^{-5}$ | 1.32×10$^{-5}$ | 280 K | 6.04×10$^{-6}$ | 7.60×10$^{-5}$ | 9.50×10$^{-6}$ |
| 300 K | 2.11×10$^{-5}$ | 1.17×10$^{-4}$ | 1.79×10$^{-5}$ | 300 K | 1.24×10$^{-5}$ | 9.79×10$^{-5}$ | 1.32×10$^{-5}$ |
| 320 K | 3.25×10$^{-5}$ | 1.50×10$^{-4}$ | 2.44×10$^{-5}$ | 320 K | 1.88×10$^{-5}$ | 1.21×10$^{-4}$ | 1.71×10$^{-5}$ |
| 340 K | 4.45×10$^{-5}$ | 1.59×10$^{-4}$ | 2.85×10$^{-5}$ | 340 K | 2.94×10$^{-5}$ | 1.43×10$^{-4}$ | 2.20×10$^{-5}$ |
| 360 K | 6.20×10$^{-5}$ | 1.62×10$^{-4}$ | 3.16×10$^{-5}$ | 360 K | 4.00×10$^{-5}$ | 1.51×10$^{-4}$ | 2.52×10$^{-5}$ |

In Chandler's original work, the initial application focused on the carbon tetrachloride (CCl$_4$) with a temperature and density independent $A_D$ value of 0.54.[39] Subsequent experimental studies confirmed the absence of temperature depencency for CCl$_4$, CF$_4$, CHF$_3$, CFCl$_3$, and CF$_3$Cl.[42, 83] However, it has been observed that molecules with large dipole moments experience a significant decrease in intermolecular interactions at higher temperatures, indicating a general decrease $A_D$.[42] This observation led to a more systematic treatment reported by Easteal and Woolf for simple polyatomic fluids.[84, 85] Motivated by these studies, we propose that the roughness parameter for water should also exhibit a temperature dependency. While a direct examination of water is challenging due to the lack of comprehensive experiments at normal density conditions[86, 87] (except for NMR studies of water at compressed supercritical states[88, 89]), it is worth noting that water exhibits strong hydrogen bonding in its liquid phase. Therefore, the strong hydrogen bonding in water suggests a temperature-dependent $A_D$ behavior, which differs from Chandler's original model.[39]

### 3-6. Recovered Diffusion Coefficients: Temperature-Dependent $A_D^{H_2O}(T)$

As shown in Table 2, the recovered CG diffusion coefficient, obtained using the temperature-independent $A_D^{H_2O} = 0.798$, strongly indicates that $A_D^{H_2O}$ should decrease at higher temperatures, such as 360 K. Furthermore, the relative accuracy of the recovered diffusion coefficients becomes more pronounced at lower temperatures. For example, as observed in Table B1, the averaged error ratio $D_{\text{CG}}^{\text{corr}}/D_{\text{FG}}$ consistently decreases with increasing temperature from 1.18 at 280 K to 0.90 at 300 K, then 0.77 at 320 K, 0.67 at 340 K, and finally 0.64 at 360 K.

In order to accurately account for the temeperature-dependent dipole moment of water, we adopt the approach proposed by Speedy et al. to introduce weak temperature dependence to the coupling parameter.[90] Based on the Arrhenius-like temperature dependence of $D$ in several liquids, Ref. 90 attributed this temperature-dependent coupling behavior to the attractive part of the intermolecular potential. Speedy originally derived the temperature dependence based on Lennard-Jones fluids as

$$D_{\text{LJ}} = D_{\text{HS}}(\sigma_B) \exp(-\epsilon/2k_B T),  \quad (30)$$

where $\sigma_B$ represents the hard sphere diameter from Boltzmann's definition[38] and $\epsilon$ corresponds to the Lennard-Jones potential energy minimum, which is the attractive part of the Lennard-Jones interaction. For Eq. (30), while the original Ref. 90 did not explicitly refer to the term $\exp(-\epsilon/2k_B T)$ as the translation-rotation coupling factor $A_D$, we note that this term is essentially the temperature-dependent coupling to the translational diffusion. Therefore, we extend Eq. (30) to the bottom-up CG models of liquids as:

$$D_{\text{CG}} = D_{\text{SHS}} \exp(-\epsilon_{\text{CG}}/2k_B T). \quad (31)$$

In contrast to $\epsilon$ in Eq. (30), $\epsilon_{\text{CG}}$ in Eq. (31) represents the CG interaction strength, which deviates from a pure hard sphere repulsion. Since bottom-up CG interactions are many-body CG PMFs, determining $\epsilon_{\text{CG}}$ for CG models is not as straightforward as in the Lennard-Jones cases.[3, 67-75] Nevertheless, as a zeroth-order approximation, we can estimate $\epsilon_{\text{CG}}$ by considering the interaction strength (i.e., potential energy minimum) within the first coordination shell of water. This approximation effectively captures the attractive interaction nature of water. As a future direction, we plan to further refine this approximation by leveraging our previous work to investigate the low temperature anormaly of CG water.[8] In Ref. 8, we developed a systematic mapping of the CG water interaction $U(R)$ to the ramp interaction $\epsilon_R$ using an energy-conserving mapping

$$4\pi\rho \int_{R_1}^{R_2} [U(R) - \epsilon_R] R^2 g(R) dR = 0, \quad (32)$$

where $R_1$ and $R_2$ correspond to the minimum and maximum pair distances, respectively, satisfying $U(R_1) = U(R_2) = \epsilon_R$.

Employing the zeroth-approximation, we extracted the $\epsilon_{\text{CG}}$ values for various force fields and temperature conditions, which are listed in Table C1, and the temperature-dependence of the roughness parameter can be accounted by estimating $\exp(-\epsilon_{\text{CG}}/2k_B T)$ factor. Remarkably, the estimated value of $\exp(-\epsilon_{\text{CG}}/2k_B T)$ at 280 K closely aligns with the temperature-independent $A_D^{\text{H}_2\text{O}}$ value determined by the acentric factor. In particular, the estimated values were 0.800 for SPC/E, 0.849 for SPC/Fw, 0.781 for TIP4P/2005, and 0.737 for TIP4P/Ice. This agreement can be interpreted as indicating that $A_D^{\text{H}_2\text{O}}$ captures the translation-rotation coupling near the freezing temperature, and it is expected to increase as temperature rises. With this factor in mind, we performed a temeperature-dependent correction, as presented in Table 3 and Fig. 8.

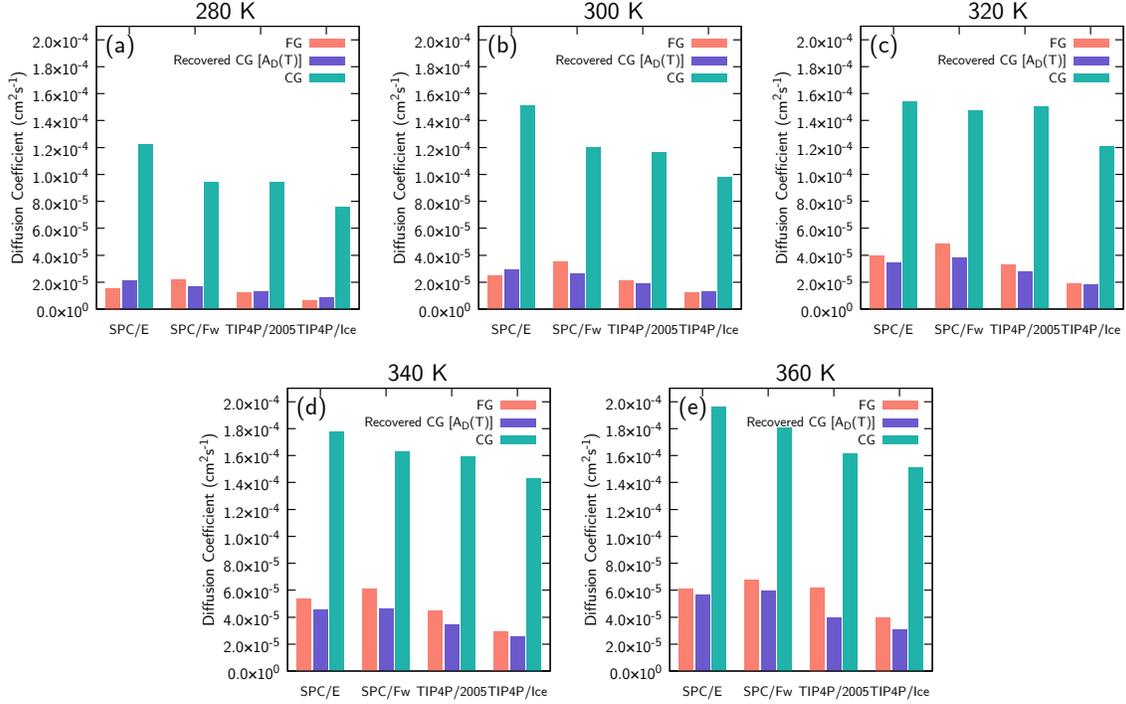

**Figure 8.** Diffusion coefficients of water for the FG (red solid) and CG (green solid) systems at different temperatures: (a) 280 K, (b) 300 K, (c) 320 K, (d) 340 K, and (e) 360 K. We recovered the FG dynamics (purple solid) from the CG diffusion coefficient values using a temperature-dependent parameter $\exp(-\epsilon_{CG}/2k_B T)$ [Eq. (31)].

Comparing the results in Table 2, we can observe a significant improvement in $D_{CG}^{corr}$ at higher temperatures. Overall, Eq. (31) can recover 92% of the FG dynamics across different force fields and temperatures, as shown in Table B2. Considering the magnitude of the accelerated diffusion $D_{CG}/D_{FG}$ ranging from 2 times to 13 times (with an average value of 5), our approach demonstrates surprisingly good performance. Given that various degrees of freedom are integrated out at the CG resolution, our approach suggests that back-mapping of dynamic information is possible under the excess entropy scaling relationship. Even though the hard sphere theory has lost its popularity due to the many developments in computer simulations, our findings suggest that physical principles extending from the hard sphere theory can be still very useful for understanding how CG dynamics corresponds to FG dynamics in molecular liquids.

**Table 3:** Diffusion coefficients of water evaluated for FG and CG models. Based on Fig. 8, the recovered CG diffusion coefficients ($D_{CG}^{corr}$) are listed for comparison, resulting in good agreement with the FG diffusion coefficient at all temperatures studied in this work (280-360 K). We demonstrated the fidelity of our approach by using the four FG force models and their corresponding BUMPer CG models: (a) SPC/E, (b) SPC/Fw, (c) TIP4P/2005, and (d) TIP4P/Ice.

| (a) SPC/E (Diffusion in cm²·s⁻¹) | | | | (b) SPC/Fw (Diffusion in cm²·s⁻¹) | | | |
| --- | --- | --- | --- | --- | --- | --- | --- |
| Temp | $D_{FG}$ | $D_{CG}$ | $D_{CG}^{corr}$ | Temp | $D_{FG}$ | $D_{CG}$ | $D_{CG}^{corr}$ |
| 280 K | 1.55×10⁻⁵ | 1.22×10⁻⁴ | 2.08×10⁻⁵ | 280 K | 2.20×10⁻⁵ | 9.39×10⁻⁵ | 1.71×10⁻⁵ |
| 300 K | 2.51×10⁻⁵ | 1.51×10⁻⁴ | 2.94×10⁻⁵ | 300 K | 3.53×10⁻⁵ | 1.20×10⁻⁴ | 2.61×10⁻⁵ |
| 320 K | 3.98×10⁻⁵ | 1.54×10⁻⁴ | 3.47×10⁻⁵ | 320 K | 4.84×10⁻⁵ | 1.47×10⁻⁴ | 3.83×10⁻⁵ |
| 340 K | 5.33×10⁻⁵ | 1.78×10⁻⁴ | 4.52×10⁻⁵ | 340 K | 6.08×10⁻⁵ | 1.63×10⁻⁴ | 4.64×10⁻⁵ |
| 360 K | 6.12×10⁻⁵ | 1.96×10⁻⁴ | 5.64×10⁻⁵ | 360 K | 6.77×10⁻⁵ | 1.81×10⁻⁴ | 5.94×10⁻⁵ |
| (c) TIP4P/2005 (Diffusion in cm²·s⁻¹) | | | | (d) TIP4P/Ice (Diffusion in cm²·s⁻¹) | | | |
| Temp | $D_{FG}$ | $D_{CG}$ | $D_{CG}^{corr}$ | Temp | $D_{FG}$ | $D_{CG}$ | $D_{CG}^{corr}$ |
| 280 K | 1.20×10⁻⁵ | 9.41×10⁻⁵ | 1.30×10⁻⁵ | 280 K | 6.04×10⁻⁶ | 7.60×10⁻⁵ | 8.77×10⁻⁶ |
| 300 K | 2.11×10⁻⁵ | 1.17×10⁻⁴ | 1.88×10⁻⁵ | 300 K | 1.24×10⁻⁵ | 9.79×10⁻⁵ | 1.31×10⁻⁵ |
| 320 K | 3.25×10⁻⁵ | 1.50×10⁻⁴ | 2.75×10⁻⁵ | 320 K | 1.88×10⁻⁵ | 1.21×10⁻⁴ | 1.84×10⁻⁵ |
| 340 K | 4.45×10⁻⁵ | 1.59×10⁻⁴ | 3.43×10⁻⁵ | 340 K | 2.94×10⁻⁵ | 1.43×10⁻⁴ | 2.55×10⁻⁵ |
| 360 K | 6.20×10⁻⁵ | 1.62×10⁻⁴ | 3.98×10⁻⁵ | 360 K | 4.00×10⁻⁵ | 1.51×10⁻⁴ | 3.07×10⁻⁵ |

**IV. Conclusions**

In this paper, we establish a comprehensive correspondence between the fine-grained (FG) and coarse-grained (CG) dynamics of water, elucidating the physical nature underlying the accelerated CG dynamics of molecular liquids. Based on the excess entropy scaling relationship of the FG and CG dynamics proposed in Paper I,[1] this work fills the gap between the entropy-free diffusion coefficients, $D_0$, and the excess entropy, $s_{ex}$, of FG and CG systems. While the differences between the FG and CG configurational entropies can be resolved at the single-site resolution, the central idea is to extract the missing rotational motion from the FG trajectory to recover the full dynamics. Given that there seems to be no rigorous Rosenfeld-like macroscopic scaling scheme for the rotational diffusion coefficient, we developed here a method to assess the translational component from such rotations, assuming that these motions only occur on the effective surface of a molecule. This idea is built upon our observation from Paper II that the CG particle can be described as an effective hard sphere, as the CG diffusion can be thought of as a hard sphere process.[2] Applying the Rosenfeld scaling to the translational component from rotations, we discover that the scaling exponent is evidently invariant under different motions, suggesting its universality.

Based on the universality of excess entropy scaling, we propose a general description of the corresponding FG and CG diffusion coefficients. We claim that the translation at the FG level contains not only the translational motion observed in the CG simulation but also the rotational motion conditioned to molecular translation. Hence, the entropy-free component of the FG diffusion coefficient should consider both rotational and translational contributions, whereas the CG dynamics from the CG simulation exhibit only translational motion due to the missing degrees of freedom. By incorporating the translation-component from rotation at the FG level to the center-of-mass translation at the CG level, we find that the full FG dynamics can be faithfully recovered from the CG dynamics when the translation-rotation coupling is also addressed.

In Paper II, we demonstrated that pure translational motion of CG particles can be described by the smooth hard sphere diffusion using the Enskog theory.[2] However, for a non-spherical system with changes in angular and linear momentum upon collision, the assumption of the Enskog theory is violated, and an effective coupling parameter should be introduced. Here, in this Paper III, based on Chandler's early observations,[31, 32, 39] we evaluate the translation-rotation coupling parameter for water from its non-sphericity described by the theorem of corresponding states. However, we observe that Chandler's treatment may not be applicable for water at different temperatures due to the presence of strong hydrogen bonding. To address this, a temperature-dependent description of the coupling is proposed by generalizing the Speedy's rough Lennard-Jones model to bottom-up CG models.[90] Consequently, an effective reduction of the final recovered diffusion coefficients notably recapitulates the reference FG diffusion coefficients at various temperatures, which was previously considered as almost impossible to understand in other systematic CG theories. It should be also noted that our framework is not confined to only water but is a generalized framework that can in principle encompass any liquid using the acentric factor or mapping to the Lennard-Jones liquid. In order to validate the fidelity of our proposed approach, future work will focus on extending this method to a variety of (bottom-up) CG liquids with diverse interaction profiles.[91] This includes exploring the behavior of liquids such as $CCl_4$ that was originally studied in Chandler's paper.[39]

One possible future direction is to apply the developed framework to glassy dynamics from the CG glass-forming liquid.[92-94] It has been demonstrated that the ratio between the translational and rotational diffusion represented by the Stokes-Einstein and Stokes-Einstein-Debye relations breaks down in the supercooled regime.[76, 95] As an extension of the recent work on the low-temperature behavior of the CG BUMPer model,[8] an examination of such dynamical heterogeneities of water could be of great interest as well. In addition, taken one step further, recovering FG dynamics from the CG dynamics of polymers can help design high-fidelity CG models since polymers exhibit highly correlated phenomena, whereas conventional CG models have yet to completely recapitulate the many-body correlations with correct dynamics.[96-100]

To summarize, our findings elucidate the differences between the CG and FG dynamics from the perspective of missing motions and entropies in the CG model. Combining this with our series of previous papers,[1, 2] our systematic approaches allow us to predict the accelerated CG diffusion coefficients solely based on FG information and also to recover the reduced FG diffusion coefficients from the accelerated CG information. We note that the single-site CG mapping using center-of-mass provides simple entropic terms and tractable diffusion phenomena, allowing us to elucidate the forward (FG → CG) and backward (CG → FG) correspondences. For the case where the CG model is no longer at a single-site resolution, a more sophisticated treatment should be developed. A resolution-based systematic description can help to understand the speed-up factors in inhomogeneous or more complex biomolecular systems, in which accurate dynamics are required to obtain correct kinetic information. An alternative yet very important future direction is a bottom-up derivation of the excess entropy scaling relationship. There has been continuous effort to derive a rigorous theory of excess entropy scaling, ranging from mode coupling theory[101] to mean first passage time[102] to Boltzmann's formula for simple deterministic Hamiltonian systems.[103] Recently, a notable contribution involved deriving this scaling relationship as a general inequality between entropy and kinetics with an exponent of $2/d$ ($d$: dimension).[104] Combining these approaches, the ultimate goal of the bottom-up CG dynamics would be to rigorously derive the scaling relationship for the many-body CG Hamiltonian by extending the Boltzmann's formula or by examining whether the general inequality yields an accurate bound for practical molecular CG systems.

**ACKNOWLEDGMENTS**

This material is based upon work supported by the National Science Foundation (NSF Grant CHE-2102677). Simulations were performed using computing resources provided by the University of Chicago Research Computing Center (RCC). J.J. acknowledges an Arnold O. Beckman Postdoctoral Fellowship from the Arnold and Mabel Beckman Foundation.

**DATA AVAILABILITY**

The data that support the findings of this work are available from the corresponding author upon request.

**APPENDIX**
**A. Estimation of hard sphere diameter of water using Barker-Henderson theory**

**Table A1:** Estimated hard sphere diameter (in Å) of CG water models at various conditions, calculated using the Barker-Henderson criterion [Eq. (25)].

| Temperature | Barker-Henderson Hard Sphere Diameter (Å) | | | |
|---|---|---|---|---|
| | SPC/E | SPC/Fw | TIP4P/2005 | TIP4P/Ice |
| 280 K | 2.534 | 2.532 | 2.532 | 2.540 |
| 300 K† | 2.542 | 2.549 | 2.563 | 2.552 |
| 320 K | 2.557 | 2.557 | 2.560 | 2.568 |
| 340 K | 2.570 | 2.567 | 2.577 | 2.583 |
| 360 K | 2.589 | 2.587 | 2.592 | 2.600 |

†: Previously reported in Ref. 2.

**B. Error ratio between the recovered FG dynamics and the reference FG dynamics**
**1. Temperature-independent case**

**Table B1:** Error ratio $D_{CG}^{corr}/D_{FG}$ between the recovered CG diffusion coefficients from this work, using Chandler's approach (Ref. 39), and the reference FG diffusion coefficients at different temperatures and FG force fields.

| Temperature | Diffusion ratio $D_{CG}^{corr}/D_{FG}$ | | | |
|---|---|---|---|---|
| | SPC/E | SPC/Fw | TIP4P/2005 | TIP4P/Ice |
| 280 K | 1.26 | 0.77 | 1.10 | 1.57 |
| 300 K | 1.05 | 0.66 | 0.85 | 1.06 |
| 320 K | 0.73 | 0.67 | 0.75 | 0.91 |
| 340 K | 0.68 | 0.62 | 0.64 | 0.75 |
| 360 K | 0.72 | 0.69 | 0.51 | 0.63 |

**2. Temperature-dependent case**

**Table B2:** Error ratio $D_{CG}^{corr}/D_{FG}$ between the recovered CG diffusion coefficients from this work, using Speedy's approach (Ref. 90), and the reference FG diffusion coefficients at different temperatures and FG force fields.

| Temperature | Diffusion ratio $D_{CG}^{corr}/D_{FG}$ | | | |
|---|---|---|---|---|
| | SPC/E | SPC/Fw | TIP4P/2005 | TIP4P/Ice |
| 280 K | 1.34 | 0.78 | 1.08 | 1.45 |
| 300 K | 1.18 | 0.74 | 0.89 | 1.06 |
| 320 K | 0.87 | 0.79 | 0.85 | 0.98 |
| 340 K | 0.85 | 0.76 | 0.77 | 0.87 |
| 360 K | 0.92 | 0.88 | 0.64 | 0.77 |

## C. Interaction Strength for CG Water

**Table C1:** $\epsilon_{CG}$ interaction parameter in Eq. (31) estimated from the BUMPer CG interactions. The determined $\epsilon_{CG}$ value corresponds to the minimum potential value in the first coordination shell [$2.5 < R < 3$ (Å)]. The temperature-dependent coupling parameter was then estimated as $\exp(-\epsilon_{CG}/2k_BT)$.

| Temperature | $\epsilon_{CG}$ from CG interactions (kcal/mol) | | | |
|---|---|---|---|---|
| | SPC/E | SPC/Fw | TIP4P/2005 | TIP4P/Ice |
| 280 K | 0.182 | 0.248 | 0.274 | 0.340 |
| 300 K | 0.139 | 0.124 | 0.209 | 0.274 |
| 320 K | 0.065 | 0.083 | 0.133 | 0.194 |
| 340 K | 0.010 | 0.026 | 0.057 | 0.107 |
| 360 K | -0.041 | -0.029 | -0.007 | 0.043 |